\documentclass[preprint,floats,showpacs]{revtex4-1}

\usepackage{amsmath}
\usepackage{amsfonts}

\usepackage{tikz}

\newcommand {\be}{\begin{equation}}
\newcommand {\ee} {\end{equation}}
\newcommand {\bea}{\begin{eqnarray}}
\newcommand {\eea} {\end{eqnarray}}

\begin{document}

%@twocolumnfalse\endcsname

\title{Free Energy Difference Fluctuations in Short-Range Spin Glasses}

\author{C.M.~Newman$^{1,2}$ and D.L. Stein$^{1,2,3,4,5}$}
\affiliation{$^1$Courant Institute of Mathematical Sciences, New York University, New York, NY 10012, 
USA\\
$^2$NYU-ECNU Institute of Mathematical Sciences at NYU Shanghai, 3663 Zhongshan Road North, Shanghai, 
200062, China\\
%$^3$Department of Physics, Yale University, P.O. Box 208120, New Haven, Connecticut 06520-8120, USA\\
%$^4$Department of Applied Physics, Yale University, P.O. Box 208284, New Haven, Connecticut 06520-8284, 
%USA\\
$^3$Department of Physics, New York University, New York, NY 10012, USA\\
$^4$NYU-ECNU Institute of Physics at NYU Shanghai, 3663 Zhongshan Road North, Shanghai, 200062, 
China\\
$^5$Santa Fe Institute, 1399 Hyde Park Rd., Santa Fe, NM 87501, USA}

%\date{January 17, 2023}

\begin{abstract}
It is generally believed (but not yet proved) that Ising spin glasses with nearest-neighbor interactions have a phase transition in three and higher dimensions to a low-temperature spin glass
phase, but the nature of this phase remains controversial, especially whether it is characterized by multiple incongruent Gibbs states.
Of particular relevance to this question is the behavior of the typical free energy difference restricted to a finite volume between two such putative Gibbs states, as well as the nature of the fluctuations of their free energy difference as the couplings within the volume vary.  In this paper we investigate these free energy difference fluctuations by introducing a new kind of metastate which classifies Gibbs states through their edge overlap values with a reference Gibbs state randomly chosen from the support of the periodic boundary condition~(PBC) metastate. We find that the free energy difference between any two incongruent pure states, regardless of the details of how they're organized into mixed states within the PBC~metastate, converges to a Gaussian (or Gaussian-like) distribution whose variance scales with the volume, proving a decades-old conjecture of Fisher and Huse. The same conclusion applies, though with some additional restrictions, to both mixed Gibbs states and ground states. We discuss some implications of these results.

%We investigate scenarios in which the low-temperature phase of short-range spin glasses comprises thermodynamic states which are
%nontrivial mixtures of multiple incongruent pure state pairs. We construct a new kind of metastate supported on Gibbs states whose edge
%overlap values with a reference state fall within a specified range. Using this metastate we show that, in any dimension, the variance of free energy difference fluctuations between pure states within a single mixed Gibbs state with multiple edge overlap values diverges linearly with the volume. This conclusion would be avoided if the distribution of edge overlap values
%in any mixed Gibbs state has at most two values: the self-overlap and an overlap between non-spin-flip related states at a smaller value, as occurs in one-step replica symmetry breaking. 
%We discuss some implications of these results.

%For short-range spin glasses, if replica symmetry breaking occurs in any finite dimension, it can be at most 1-step RSB.
\end{abstract}

%\pacs{pacs} %]

\maketitle

\section{Introduction}
\label{sec:intro}

The thermodynamic behavior at low temperature of short-range spin glasses in finite dimension remains an active area of research, with major questions still unresolved. These include whether the spin glass phase, assuming it exists, is supported on multiple non-spin-flip related Gibbs states at positive temperature and/or ground states at zero temperature;  whether {\it individual\/} Gibbs states at positive temperature consist of a single spin-reversed pure state pair (assuming Ising spins and broken spin-flip symmetry), or instead a mixture of multiple distinct pure state pairs; and whether a small magnetic field destroys the spin glass phase in finite dimension or merely rearranges the structure of the Gibbs states.

A central question that arises if the spin glass phase comprises multiple Gibbs states is how the typical free energy difference between distinct Gibbs states scales with the volume considered; because this is a random quantity depending on the coupling realization, the fluctuations of this free energy difference are also a central quantity of interest. This issue doesn't arise if the spin glass phase comprises a single pure state pair, although in that case one can ask about the typical free energy cost of replacing a finite subset of spins of one pure state with those of its spin-reversed partner. We will not consider that question in this paper, but will focus instead on free energy differences between non-spin-reversed Gibbs states.  These fluctuations are of particular importance in determining salient spin glass properties such as frustration, stiffness, and whether such distinct thermodynamic states occur at all.  

There are two settings in which these questions can be studied: one is within the context of strictly finite systems under changes of boundary conditions, and the second is in a thermodynamic setting where one studies free energy differences in finite-volume subsets of two distinct infinite-volume Gibbs states.  Both are important for a complete understanding of spin glass behavior, but in this paper we shall consider only the second situation. We find that in general free energy differences between incongruent pure states (defined in Sect.~\ref{sec:incong}) display Gaussian-like behavior in the limit as volume goes to infinity, with variance scaling as the volume; similar results hold for entire mixed Gibbs states at positive temperature and ground states at zero temperature, but with some restrictions which are detailed in Secs.~\ref{sec:mixed}~and~\ref{sec:zero}. These results prove a version of a conjecture of Fisher and Huse~\cite{HF87,FH87} that typical free energy differences between incongruent pure states within a region of volume~$L^d$ scale as~$L^{d/2}$.

The paper is organized as follows: In Sect.~\ref{sec:finite} we introduce the Edwards-Anderson~(EA) Ising Hamiltonian~\cite{EA75}, briefly discuss some scenarios that have been proposed for its low-temperature behavior, and present some relevant results from numerical studies that apply to all pictures. In Sect.~\ref{sec:meta} we review the notion of a metastate~\cite{AW90,NS96c,NRS23b}, with particular attention to the structure of periodic boundary condition metastates for the EA Ising Hamiltonian under the different scenarios of interest. In Sect.~\ref{sec:overlap} we present a result that is central to what follows, namely that the edge (or spin) overlap between two infinite-volume Gibbs states is invariant under finite changes of couplings.  In Sect.~\ref{sec:incong} we discuss the notion of Gibbs state~incongruence~\cite{HF87,FH87} and present some results related to it.  In Sect.~\ref{sec:nontrivial} we review rigorous results on properties of a metastate supported on Gibbs states which are nontrivial mixtures of multiple pure state pairs. Of particular interest here is the metastate barycenter, i.e., the Gibbs state that results when the metastate is averaged over all states in its support. In Sect.~\ref{sec:restricted} we introduce the restricted metastate, a new type of metastate indexed by a reference Gibbs state chosen randomly from a periodic boundary condition metastate, and which classifies Gibbs states according to their edge overlap with the reference state.  In Sect.~\ref{sec:metainc} we introduce the notion of metastate incongruence and prove that restricted metastates satisfy this property. In Sect.~\ref{sec:pure} we prove a portion of our main result, namely that the free energy difference between two incongruent pure states, whether chosen from the same or different mixed Gibbs states, have fluctuations (with respect to the couplings) whose variance scales with the volume.  We then extend these results to free energy differences between entire mixed Gibbs states at positive temperature~(Sect.~\ref{sec:mixed}) and ground states at zero temperature~(Sect.~\ref{sec:zero}), under the condition (consistent with replica symmetry breaking) that the overlap distribution of the metastate barycenter consists of a single $\delta$-function. In Sect.~\ref{sec:Gaussian} we show that the distribution of free energy differences in all of the above cases has a tail at least as big as a Gaussian, and use that to obtain some new results in two dimensions. We conclude in Sect.~\ref{sec:discussion} with a summary and discussion. 

A preliminary report of this study confined to pure states from the same mixed Gibbs state appeared as an informal note in~\cite{NS24}.

\vfill\eject

\section{Spin glass phase in finite dimension}
\label{sec:finite}

A canonical model for studying the behavior of realistic spin glasses is the well-known Edwards-Anderson Ising model~\cite{EA75}, whose Hamiltonian in zero magnetic field is
\begin{equation}
\label{eq:EA}
{\cal H}_J=-\sum_{(x,y)} J_{xy} \sigma_x\sigma_y\, , 
\end{equation}
where we take the sites $x$ to be vertices of the $d$-dimensional cubic lattice~$\mathbb{Z}^d$. In~(\ref{eq:EA}) $\sigma_x=\pm 1$ is the Ising spin at site $x\in\mathbb{Z}^d$ and $(x,y)$ denotes an edge in~$\mathbb{E}^d$, the nearest-neighbor edge set of $\mathbb{Z}^d$. The couplings $J_{xy}$ are independent, identically distributed random variables chosen from a distribution $\nu(dJ_{xy})$ symmetrically distributed about zero. In this paper assumptions on~$\nu$ are modest: e.g., it suffices, as in~\cite{ANSW14}, that it be continuous with finite fourth moment. We denote by $J$ a particular realization of the couplings.

The statistical mechanics of the EA Hamiltonian~(\ref{eq:EA}) remains theoretically unresolved in dimensions greater than one.  It is generally accepted, mostly due to numerical studies, that there is no phase transition at positive temperature in two dimensions~\cite{BY86}, but that in three dimensions and above there is a thermodynamic phase transition at a dimension-dependent temperature $T_c(d)>0$~\cite{BY86,GSNLAI91,PC99,Ball00,Boettcher05}. In all dimensions, there is a unique paramagnetic phase at temperatures above $T_c(d)$ in which the thermal expectation value of any spin $\langle\sigma_x\rangle=0$ due to the global spin-flip symmetry ($\sigma_x\to-\sigma_x$ for all $x$) of the Hamiltonian~(\ref{eq:EA}), but the nature of the ``spin-glass phase" for temperatures below~$T_c(d)$ remains unresolved.  There are four proposed scenarios for the spin glass phase that as of this writing are consistent with the majority of numerical results and are also mathematically consistent: replica symmetry breaking (RSB)~\cite{Parisi79,Parisi83,MPSTV84a,MPSTV84b,MPV87,MPRR96,MPRRZ00,NS02,NS03b,Read14,NRS23b}, droplet-scaling~~\cite{Mac84,BM85,BM87,FH86,FH88b}, trivial-nontrivial spin overlap~(TNT)~\cite{MP00,PY00}, and chaotic pairs~\cite{NS96c,NS97,NSBerlin,NS03b}.

All of these pictures assume broken spin-flip symmetry below $T_c(d)$, i.e., the Edwards-Anderson order parameter
\begin{equation}
\label{eq:OP}
q_{EA}=\lim_{L\to\infty}\vert\Lambda_L\vert^{-1}\sum_{x\in\Lambda_L}\langle\sigma_x\rangle_\alpha^2
\end{equation}
is strictly positive, where $\Lambda_L=[-L,L]^d\subset\mathbb{Z}^d$, $\vert\Lambda_L\vert=(2L+1)^d$ and $\alpha$ denotes a pure state in the spin glass phase. (It was recently proved
that $q_{EA}$ has the same value for all pure states within a single mixed state~\cite{NRS23a} and for all mixed states in the metastate~\cite{Read24}; because it is translation-invariant it therefore has the same value in a.e.~coupling configuration. Hence $q_{EA}$ has no $\alpha$ or $J$ dependence.)  Due to the spin-flip symmetry of the Hamiltonian, a nonzero EA~order parameter $q_{EA}>0$ implies that every pure state, generated by an infinite sequence of spin-symmetric boundary conditions such as free or periodic, has a globally spin-reversed counterpart with equal weight. Consequently in these situations we refer to ``pure state pairs'' (or ground state pairs at zero temperature). Formally, a pure state pair $(\alpha,\overline{\alpha})$ satisfies $\langle\sigma_{x_1}\sigma_{x_2}\ldots\sigma_{x_n}\rangle_\alpha=\langle\sigma_{x_1}\sigma_{x_2}\ldots\sigma_{x_n}\rangle_{\overline{\alpha}}$ for $n$ even and $\langle\sigma_{x_1}\sigma_{x_2}\ldots\sigma_{x_n}\rangle_\alpha=-\langle\sigma_{x_1}\sigma_{x_2}\ldots\sigma_{x_n}\rangle_{\overline{\alpha}}$ for $n$ odd. For convenience of notation~$\alpha$ will henceforth be used to denote a pure state pair rather than a single pure state unless otherwise noted.

The presence of broken spin-flip symmetry in the spin glass phase is supported by numerical studies~(see, e.g.,~\cite{MPRRZ00}), for example in the observation of a pair of $\delta$-functions at $\pm q_{EA}$ in most, if not all, simulations of the overlap structure (to be discussed below) of the EA Ising spin glass at low temperature. (Much of the controversy regarding the thermodynamic structure of the spin glass phase revolves around what, if anything, lies between that pair of $\delta$-functions.) In what follows we will therefore assume that, in the EA~Ising~model in three dimensions and higher, there is a thermodynamic phase transition from a paramagnetic phase at high temperature to a spin glass phase with broken spin-flip symmetry at low temperature.

The differences among the four pictures have been described in detail elsewhere~\cite{NS03a,NS22,NRS23b} to which we refer the interested reader; below we briefly summarize those differences of relevance to the work described here. For specificity, consider a deterministic (i.e., chosen independently of couplings and boundary conditions) infinite sequence of volumes $\Lambda_L=[-L,L]^d$ centered at the origin and each with periodic boundary conditions. Above~$T_c(d)$ this sequence of finite-volume Gibbs states generated by Hamiltonian~(\ref{eq:EA}) will converge to a single pure infinite-volume Gibbs state (the paramagnetic state) in which $\langle\sigma_x\rangle=0$ for all sites~$x$.  Below $T_c(d)$, in all four pictures pure states always appear in spin-reversed pairs, but other than that the pictures differ in several ways. First is the question of convergence: if droplet-scaling or TNT hold, the sequence converges to a Gibbs state which is a symmetric mixture of a single pure state pair.  In this paper we will call such a Gibbs state a {\it trivial\/} mixed state; if the mixed state is supported on more than a single pure state pair, we will call it a {\it nontrivial\/} mixed state.  (The major difference between droplet-scaling and TNT lies in the nature of their low-energy, large-lengthscale excitations, but that is not relevant to the present discussion.)

If either RSB or chaotic pairs hold, the finite-volume Gibbs states generated by a deterministic sequence of volumes will {\it not\/} converge~\cite{NS92} to a single limit, but there will instead be two or more subsequences converging to different mixed Gibbs states. (We will further refine this in Sect.~\ref{sec:incong}.) In the chaotic pairs picture, each of these limiting Gibbs states is a trivial mixture of a single pure state pair, but in RSB the mixture is nontrivial: each Gibbs state generated through this process is predicted to be a mixture of a countable infinity of pure state pairs. So while RSB and chaotic pairs both predict multiple Gibbs states in the spin glass phase, of the four pictures presented here only RSB predicts the existence of nontrivial mixed Gibbs states.

\section{Metastates}
\label{sec:meta}

Two of the four pictures discussed above assume the existence of multiple pure state pairs in the spin glass phase. The resulting lack of convergence of finite-volume Gibbs states along a deterministic sequence of volumes then presents an obstacle to analyzing the thermodynamics of the EA Ising spin glass at low temperature. To circumvent this difficulty one may consider instead the {\it distribution\/} of thermodynamic states along the sequence. 

An analogy with the behavior of chaotic dynamical systems is instructive here~\cite{NS96c}: although the behavior of a chaotic system is deterministic but (effectively) unpredictable, the {\it fraction\/} of time spent by the particle in different regions of phase space (roughly speaking) does converge to some distribution~$\kappa$. In a similar manner it can be shown that (again roughly speaking) the fractions of $\Lambda_L$'s with convergence to different limiting Gibbs states also determine some distribution~$\kappa_J$.  Just as an individual thermodynamic state is a probability measure on spin configurations, $\kappa_J$ is a probability measure on the thermodynamic states themselves. As the notation indicates, the metastate $\kappa_J$ depends on the coupling realization~$J$, but also depends on the sequence of boundary conditions chosen to generate the thermodynamic state(s). ($\kappa_J$ also clearly depends on both $\beta$ and $d$, but our notation omits this dependence.) We therefore refer to periodic boundary condition metastates (all volumes along the sequence are assigned periodic boundary conditions), free boundary condition metastates, and so on. 

There are two known constructions (Aizenman-Wehr (AW)~\cite{AW90} and Newman-Stein (NS)~\cite{NS96c}; see~\cite{NRS23b} for a review) that result in a metastate; in what follows either construction can be used, given that there are always subsequences of volumes along which the two approaches converge to the same metastate~\cite{NSBerlin}. 

Here we provide a brief summary of the definition of a metastate: let $\Sigma=\{-1,+1\}^{\mathbb{Z}^d}$ be the set of all infinite-volume Ising spin configurations and let ${\cal M}_1(\Sigma)$ be the set of (regular Borel) probability measures on~$\Sigma$. A metastate $\kappa_J$ at a given inverse temperature~$\beta$ is a measurable mapping $J\mapsto\kappa_J$ from $\mathbb{R}^{{\mathbb E}^d}$ to $\{\text{probability measures on}\ {\cal M}_1(\Sigma)\}$, with the following properties~\cite{ANSW14,NRS23b}:

\medskip

{\bf 1. Support on Gibbs states.} Let the set of Gibbs states corresponding to the coupling realization $J$ (at a given~$\beta$) be denoted by ${\mathcal G}_J$.  Then every state sampled from $\kappa_J$ is a thermodynamic (Gibbs) state for the realization $J$:
\begin{equation}
\label{eq:1}
\kappa_{J}\Bigl(\mathcal G_{J}\Bigr)=1.  %\ \text{$\nu$-a.s.}\nonumber
\end{equation}

{\bf 2. Coupling Covariance.} For  $B\subset \mathbb{Z}^d$ finite, $J_B\in \mathbb{R}^{\mathbb{E}(B)}$ (where $\mathbb{E}(B)$ is the set of edges in $B$), and $\Gamma$ a Gibbs state, we define the operation ${\cal L}_{J_B} : \Gamma \mapsto {\cal L}_{J_B}\Gamma$ on~${\cal M}_1(\Sigma)$ by its effect on the expectation $\langle\cdots\rangle_\Gamma$ 
in $\Gamma$,
\begin{equation}
\label{eq: gamma L}
 \left\langle f(\sigma)\right\rangle_{{\cal L}_{J_B}\Gamma}= \frac{\left\langle f(\sigma) \exp\Bigl(-\beta H_{J_B}(\sigma)\Bigr)\right\rangle_\Gamma}
{\left\langle\exp\Bigl(-\beta H_{J_B}(\sigma)\Bigr)\right\rangle_\Gamma}\, ,
\end{equation}
%Note that the coupling modification is simply a change of density. In particular, it is easy to check that $\Gamma$ and $L_{\Delta J}\Gamma$ 
%are equivalent.
which describes the effect of modifying the couplings within $B$. We require that the metastate 
be covariant under local modifications of the couplings, i.e., for any measurable subset $A$ of  $\mathcal M_1(\Sigma)$,

\begin{equation}
\label{eq:2}
\kappa_{J+J_B}(A)= \kappa_{J}({\cal L}_{J_B}^{-1}A)
\end{equation}
where ${\cal L}_{J_B}^{-1} A=\Bigl\{\Gamma\in \mathcal M_1(\Sigma): {\cal L}_{J_B}\Gamma \in A\Bigr\}$.
%\end{df}

\medskip

{\bf 3. Translation Covariance.} For any lattice translation $\tau$ of $\mathbb{Z}^d$ and any measurable subset $A$ of  
$\mathcal M_1(\Sigma)$,
\begin{equation}
\label{eq:3}
\kappa_{\tau J}(A)=\kappa_{J}(\tau^{-1}A).
\end{equation}
%\smallskip 
%\end{enumerate}
i.e., a uniform lattice shift does not affect the metastate properties. This is guaranteed when one constructs a metastate using periodic boundary conditions to generate all finite-volume Gibbs states; in the infinite-volume limit, the Gibbs states (and therefore the metastate) will inherit the torus-translation covariance of the finite-volume Gibbs states. Henceforth $\kappa_J$ shall refer only to a PBC metastate of the EA~Hamiltonian~(\ref{eq:EA}).

We now turn to the PBC~metastate structure of the various scenarios for the spin glass phase. At temperatures above~$T_c$, with any deterministic boundary condition, the metastate comprises a single pure state, namely the paramagnetic state. At temperatures below~$T_c$, the situation is as follows: for both two-state pictures (droplet-scaling and TNT) PBC~metastates live on a single mixed Gibbs state, which is a trivial mixture of a single spin-reversed pure state pair. In the chaotic pairs picture, PBC metastates are dispersed; such a metastate is supported on two or more trivial mixed Gibbs states. 

All of these metastates have a simple structure. However, metastates within the RSB picture (or any picture with nontrivial mixed Gibbs states) are more complicated;  before turning to those, we need to introduce some additional concepts and results.

\section{Overlap invariance}
\label{sec:overlap}

Let $E_L = \mathbb{E}(\Lambda_L)$, the edge set within $\Lambda_L$, and define the edge (or bond) overlap between two states $\alpha$ and $\beta$ as
\begin{equation}
\label{eq:overlap}
q^{(e)}_{\alpha\beta}=\overline{\langle\sigma_x\sigma_y\rangle_\alpha\langle\sigma_x\sigma_y\rangle_\beta}=\lim_{L\to\infty}\frac{1}{d\vert\Lambda_L\vert}\sum_{\langle xy\rangle\in E_L}\langle\sigma_x\sigma_y\rangle_\alpha\langle\sigma_x\sigma_y\rangle_\beta\, .
\end{equation}
where $\overline{f(\sigma_x\sigma_y)}=\lim_{L\to\infty}\frac{1}{d\vert\Lambda_L\vert}\sum_{\langle xy\rangle\in E_L} f(\sigma_x\sigma_y)$ denotes the spatial average of a measurable, bounded function~$f$ of edge variables.  (The limit on the RHS of~(\ref{eq:overlap}) is expected to exist, but if not, it can be replaced by the lim sup, which is guaranteed to exist~\cite{Royden10}. For simplicity, we will assume throughout that the limit in~(\ref{eq:overlap}) exists for all pairs of pure states in the metastate.)
 
We begin by proving the following Lemma:

\medskip

{\bf Lemma 4.1.} For a.e.~coupling realization~$J$ and fixed inverse temperature~$\beta<\infty$, the bond overlap $q^{(e)}_{\alpha\beta}$ is invariant under a finite change in couplings (i.e., finite changes in the values of finitely many couplings).
\medskip

{\bf Proof.} It is sufficient to consider how the nearest-neighbor correlation function~$\langle\sigma_x\sigma_y\rangle_\alpha$, evaluated in the pure state $\alpha$, is affected by a change in an arbitrary coupling $J_{uv}\to J_{uv}+\Delta$, where $\Delta$ is any finite real number. As noted above, such a transformation maps the pure state $\alpha$ to a pure state $\alpha'$~\cite{AW90,NS96c}, and, by direct calculation,
 $\langle\sigma_x\sigma_y\rangle_\alpha$ is transformed according to
 \begin{equation}
 \label{eq:transform}
 \langle\sigma_x\sigma_y\rangle_\alpha\to\langle\sigma_x\sigma_y\rangle_{\alpha'}=\frac{\langle\sigma_x\sigma_y\rangle_\alpha+\tanh(\beta\Delta)\langle\sigma_x\sigma_y\sigma_u\sigma_v\rangle_\alpha}{1+ \tanh(\beta\Delta)\langle\sigma_u\sigma_v\rangle_\alpha}\, .
 \end{equation}
 Because $\alpha$ is a pure state, it satisfies a space-clustering property~\cite{vEvH84}:  given two localized events~$A$ and $B$ in the configuration space and writing $\|x\|$ for the Euclidean length of $x$, 
 \begin{equation}
 \label{eq:clustering1}
 \lim_{\| x\|\to\infty}\vert\langle A\tau_x B\rangle_\alpha-\langle A\rangle_\alpha\langle\tau_x B\rangle_\alpha\vert=0\, ;
 \end{equation}
so if $0'$ is a neighbor of the origin~$0$,
\begin{equation}
 \label{eq:clustering2}
 \lim_{\| x\|\to\infty}\vert\langle\sigma_u\sigma_v\tau_x(\sigma_0\sigma_{0'})\rangle_\alpha-\langle\sigma_u\sigma_v\rangle_\alpha\langle\tau_x(\sigma_0\sigma_{0'})\rangle_\alpha\vert=0\, .
 \end{equation}
 The clustering condition~(\ref{eq:clustering2}) for a pure state $\alpha$ can be rewritten with $D_{xy}$ the Euclidean distance between the origin and the edge $(x,y)$,
 \begin{equation}
 \label{eq:clustering}
 \lim_{K\to\infty}\sup_{D_{xy}>K}\vert\langle \sigma_u\sigma_v\sigma_x\sigma_y\rangle_\alpha-\langle \sigma_u\sigma_v\rangle_\alpha\langle\sigma_x\sigma_y\rangle_\alpha\vert=0\, .
 \end{equation}
Let $F(K)=\sup_{D_{xy}>K}\vert\langle \sigma_u\sigma_v\sigma_x\sigma_y\rangle_\alpha-\langle \sigma_u\sigma_v\rangle_\alpha\langle\sigma_x\sigma_y\rangle_\alpha\vert$. Fix~$(u,v)$ and consider the spatial average 
\begin{equation}
\label{eq:avg}
\lim_{K\to\infty}\frac{1}{K^d}\sum_{\langle xy\rangle\in E_K}\vert\langle \sigma_u\sigma_v\sigma_x\sigma_y\rangle_\alpha-\langle \sigma_u\sigma_v\rangle_\alpha\langle\sigma_x\sigma_y\rangle_\alpha\vert\nonumber\, .
\end{equation}
Now consider two ($d$-dimensional) cubes, both centered at the origin, with sides $\tilde K$ and $K$ such that ${\tilde K}\ll K$. The previous sum can be split up as
\begin{eqnarray}
\label{eq:split}
\lim_{K\to\infty}\frac{1}{K^d}\Bigl[\sum_{\langle xy\rangle\in E_{\tilde K}}\vert\langle \sigma_u\sigma_v\sigma_x\sigma_y\rangle_\alpha-\langle \sigma_u\sigma_v\rangle_\alpha\langle\sigma_x\sigma_y\rangle_\alpha\vert+\nonumber\\ 
\sum_{\langle xy\rangle\in{E_K}\backslash E_{\tilde K}}\vert\langle \sigma_u\sigma_v\sigma_x\sigma_y\rangle_\alpha-\langle \sigma_u\sigma_v\rangle_\alpha\langle\sigma_x\sigma_y\rangle_\alpha\vert\Bigr]\, .\nonumber
\end{eqnarray}
For fixed ${\tilde K}$ the first term vanishes in the limit while the second term is bounded from above by~$F({\tilde K})$
% \begin{equation}
% \label{eq:bound}
% \frac{1}{K^d}\sum_{\langle xy\rangle\in{E_K}/E_{\tilde K}}\Bigl[\vert\langle \sigma_u\sigma_v\sigma_x\sigma_y\rangle_\alpha-\langle \sigma_u\sigma_v\rangle_\alpha\langle\sigma_x\sigma_y\rangle_\alpha\vert\Bigr]\le\frac{1}{K^d}F({\tilde K})(K^d)=F({\tilde K})\nonumber
% \end{equation}
which by~(\ref{eq:clustering}) goes to zero as ${\tilde K}\to\infty$.

In the remainder of the argument we will encounter 
\begin{equation}
\frac{1}{K^d}\sum_{\langle xy\rangle\in{E_K}/E_{\tilde K}}\langle\sigma_x\sigma_y\rangle_\alpha\Bigl(\langle \sigma_u\sigma_v\sigma_x\sigma_y\rangle_\alpha-\langle \sigma_u\sigma_v\rangle_\alpha\langle\sigma_x\sigma_y\rangle_\alpha\Bigr)\nonumber
\end{equation} 
and similar quantities, but given that the magnitude of any $p$-spin correlation is bounded by one for Ising spins, all such terms vanish in the limit as well. 
%$\vert\langle\sigma_x\sigma_y\rangle_\alpha\vert\le 1$ , 
%\begin{equation}
%\label{eq:ext}
%\lim_{K\to\infty}\frac{1}{K^d}\sum_{\langle xy\rangle\in{E_K}/E_{\tilde K}}\langle\sigma_x\sigma_y\rangle_\alpha\Bigl(\langle \sigma_u\sigma_v\sigma_x\sigma_y\rangle_\alpha-\langle \sigma_u\sigma_v\rangle_\alpha\langle\sigma_x\sigma_y\rangle_\alpha\Bigr)= 0\, .
%\end{equation}

We can now evaluate the change in the edge overlap $q^{(e)}_{\alpha\beta}$ between two pure states~$\alpha$ and~$\beta$ under a finite change of couplings.  Under the 
change~$J_{uv}\to J_{uv}+\Delta$ the correlations in $\alpha$ and $\beta$ transform according to~(\ref{eq:transform}) so the transformed overlap $q^{(e)}_{\alpha'\beta'}$ is
\begin{eqnarray}
\label{eq:overlaptransform}
q^{(e)}_{\alpha'\beta'}&=&\lim_{L\to\infty}\frac{1}{dL^d}\sum_{\langle xy\rangle\in E_L}\langle\sigma_x\sigma_y\rangle_{\alpha'}\langle\sigma_x\sigma_y\rangle_{\beta'}
=\frac{1}{\Bigl(1+\tanh(\beta\Delta)\langle\sigma_u\sigma_v\rangle_\alpha\Bigr)\Bigl(1+\tanh(\beta\Delta)\langle\sigma_u\sigma_v\rangle_\beta\Bigr)}\nonumber\\
&\times&\lim_{L\to\infty}\frac{1}{dL^d}\sum_{\langle xy\rangle\in E_L}\Bigl(\langle\sigma_x\sigma_y\rangle_\alpha+\tanh(\beta\Delta)\langle\sigma_x\sigma_y\sigma_u\sigma_v\rangle_\alpha\Bigr)\nonumber\\
&\times&\Bigl(\langle\sigma_x\sigma_y\rangle_\beta+\tanh(\beta\Delta)\langle\sigma_x\sigma_y\sigma_u\sigma_v\rangle_\beta\Bigr)=\lim_{L\to\infty}\frac{1}{dL^d}\sum_{\langle xy\rangle\in E_L}\langle\sigma_x\sigma_y\rangle_{\alpha}\langle\sigma_x\sigma_y\rangle_{\beta}=q^{(e)}_{\alpha\beta}\, .
\end{eqnarray}
This completes the argument.  $\diamond$

\medskip

{\it Remark.\/} Although the statement of~Lemma~4.1 was restricted to the bond overlap, the conclusion holds also for the usual spin overlap $q_{\alpha\beta}=\lim_{L\to\infty}\frac{1}{L^d}\sum_{x\in\Lambda_L}\langle\sigma_x\rangle_\alpha\langle\sigma_x\rangle_\beta$.

\medskip

\section{Incongruence between thermodynamic states}
\label{sec:incong}

The property of incongruence~\cite{HF87,FH87}  is central to a discussion concerning multiplicity of pure states in Ising spin glasses.  We use the following definition from~\cite{ANSW14}: 

\medskip

{\bf Definition 5.1.}  Two (pure or mixed) thermodynamic states $\alpha$ and $\beta$ are defined to be incongruent if for some $\epsilon>0$ there is a subset of edges $(x_0,y_0)$ with strictly positive lower density such that $\vert\langle\sigma_{x_0}\sigma_{y_0}\rangle_\alpha- \langle\sigma_{x_0}\sigma_{y_0}\rangle_\beta\vert > \epsilon$. 

\medskip

Incongruence can be thought of in several different ways: for example, the definition of incongruence for two states~$\alpha$ and~$\beta$ is equivalent to the condition $q^{(e)}_{\alpha\beta}<q^{(e)}_{\alpha\alpha}$, where the self-overlap~$q^{(e)}_{\alpha\alpha}$ is the same for all pure states within a single mixed Gibbs state~\cite{NRS23a}. It can also be thought of in terms of the {\it interface\/} between spin configurations chosen from the two incongruent pure states; the interface is defined as the set of couplings satisfied in one spin configuration but not the other.  (A coupling $J_{xy}$ is said to be satisfied in a spin configuration $\sigma$ if $J_{xy}\sigma_x\sigma_y>0$.) Roughly speaking, if the pure states are incongruent, then their interface dimension $d_s$ is equal to the spatial dimension $d$ of the entire system: $d_s=d$.

For systems whose couplings have only a single sign, such as ferromagnets and antiferromagnets, incongruent pure states cannot exist. In those systems, multiple pure states can exist in certain dimensions, but these must be {\it regionally congruent\/}~\cite{HF87,FH87}: their interfaces satisfy $d-1\le d_s<d$. Roughly speaking, two regionally congruent pure states look similar (modulo a global spin flip) most everywhere on the lattice, while two incongruent pure states look dissimilar most everywhere.

One of the important ways in which a spin glass differs from these other systems is that incongruent pure states can in principle occur (and moreover their existence is central to both the RSB picture~\cite{Parisi79,Parisi83,MPSTV84a,MPSTV84b,MPV87}) and the chaotic pairs picture~\cite{NS96c,NS97,NSBerlin,NS03b}).  While at present the question of their existence in short-range models remains unresolved, the next theorem will be important in what follows:

\medskip

{\bf Theorem 5.2.}  Given the Hamiltonian~(\ref{eq:EA}) and a~$\kappa_J$ constructed from it, all non-spin-flip-related pure states in $\kappa_J$ are mutually incongruent.

\medskip

{\bf Proof.}  This was proved in~\cite{NS01c} for ground states. To extend the result to positive temperature,  we consider pure states~$\alpha$ and $\alpha'$ and use the torus-translation-covariance of the finite-volume Gibbs states leading to the construction of $\kappa_J$ to show that the triple $(J,\alpha,\alpha')$ and its distribution $\kappa^\dagger$ is covariant under all translations of the infinite-volume cubic lattice.  The translation-covariance of~$\kappa^\dagger$ allows its decomposition into components in which translation-ergodicity holds.  For any $\epsilon>0$ and each bond $\langle x,y\rangle$, consider the event $A^\epsilon_{(x,y)}$ that $\vert\langle\sigma_x\sigma_y\rangle_{\alpha}-\langle\sigma_x\sigma_y\rangle_{\alpha'}\vert>\epsilon$. In each ergodic component, either the probability of $A^\epsilon_{(x,y)}$ is zero or else it equals some $c>0$. 

Therefore the event~$A^\epsilon_{(x,y)}$ has either probability zero for every edge or else has probability~$c$ for every edge. In the latter case, by the spatial ergodic theorem the spatial density of edges such that $A^\epsilon_{(x,y)}$ occurs must equal~$c$, so $\alpha$ and $\alpha'$ are incongruent. In the former case, for any $\epsilon>0$ and each $\langle x,y\rangle$, the event  $A^\epsilon_{(x,y)}$ has probability zero. It follows that the event~$A^\epsilon=\cup_{\langle x,y\rangle\in\mathbb{E}^d}A^\epsilon_{(x,y)}$, which is the probability that one or more edges in $\mathbb{E}^d$ satisfies $\vert\langle\sigma_{x_0}\sigma_{y_0}\rangle_\alpha- \langle\sigma_{x_0}\sigma_{y_0}\rangle_{\alpha'}\vert > \epsilon$,  also has probability zero by countable additivity~\cite{Read2}.  Thus there is zero probability of a $(J,\alpha,\alpha')$ triple such that the condition $\vert\langle\sigma_x\sigma_y\rangle_\alpha-\langle\sigma_x\sigma_y\rangle_{\alpha'}\vert>\epsilon$ is satisfied only for a zero density of bonds. 

This shows that any pair $(\alpha,\alpha')$ of pure states chosen from $\kappa_J$  are either (a) incongruent or (b) $\langle\sigma_x\sigma_y\rangle_\alpha=\langle\sigma_x\sigma_y\rangle_{\alpha'}$ for all $\langle x,y\rangle\in\mathbb{E}^d$. To go further and show that $\alpha=\alpha'$ one needs to show that {\it all\/} $k$-spin ($k=1,2,3\ldots$) correlation functions are the same in $\alpha$ and $\alpha'$. To show this, assume that $\alpha$ and $\alpha'$ obey alternative~(b) above. This can be restated as $q^{(e)}_{\alpha\alpha'}=q^{(e)}_{EA}$, and by Lemma~4.1 the edge overlap is unchanged under any finite change in couplings. Therefore two pure states which satisfy alternative~(b) for a given $J$ continue to satisfy alternative~(b) under any finite change in couplings.  At fixed positive temperature the correlation functions of pure states are differentiable functions of the couplings, so given an arbitrary $\langle x,y\rangle\in\mathbb{E}^d$, one can take the derivative of $\langle\sigma_x\sigma_y\rangle$ in both $\alpha$ and $\alpha'$ with respect to $J_{yz}$, where $z$ is a nearest neighbor to $y$ but not $x$:
\begin{equation}
\label{eq:2spin}
\frac{\partial}{\partial J_{yz}}\langle\sigma_x\sigma_y\rangle_\alpha=\frac{\partial}{\partial J_{yz}}\langle\sigma_x\sigma_y\rangle_{\alpha'} 
\Rightarrow\langle\sigma_x\sigma_z\rangle_\alpha-\langle\sigma_x\sigma_y\rangle_\alpha\langle\sigma_y\sigma_z\rangle_\alpha=\langle\sigma_x\sigma_z\rangle_{\alpha'}-\langle\sigma_x\sigma_y\rangle_{\alpha'}\langle\sigma_y\sigma_z\rangle_{\alpha'}\, .
\end{equation}
Given that $\langle\sigma_x\sigma_y\rangle_{\alpha}\langle\sigma_y\sigma_z\rangle_{\alpha}=\langle\sigma_x\sigma_y\rangle_{\alpha'}\langle\sigma_y\sigma_z\rangle_{\alpha'}$, it follows that $\langle\sigma_x\sigma_z\rangle_\alpha=\langle\sigma_x\sigma_z\rangle_{\alpha'}$.  This procedure can be repeated indefinitely to show that {\it all\/} two-spin correlation functions are the same in $\alpha$ and $\alpha'$.

The same procedure can be used to show that all four-spin correlation functions are the same in $\alpha$ and $\alpha'$ by first taking the derivative of $\langle\sigma_x\sigma_y\rangle_\alpha=\langle\sigma_x\sigma_y\rangle_{\alpha'}$ with respect to $J_{uv}$, where $\langle u,v\rangle$ is any edge noncontiguous to $\langle x,y\rangle$. One can then extend this result to all four-point and higher even-$k$-spin correlations.

We now turn to $k$-spin correlations for odd~$k$, beginning with $k=1$. The clustering property~(\ref{eq:clustering}) for pure states can be rewritten as follows. Let $\sigma_0$ be the spin at the origin and as before let $\|y\|$ denote the Euclidean distance between $\sigma_y$ and $\sigma_0$. Then for any $\epsilon>0$, there exists an $R_\epsilon>0$ such that for any $\|y\|>R_\epsilon$, 
\begin{equation}
\label{eq:cluster2}
\vert\langle\sigma_0\sigma_y\rangle_\alpha-\langle\sigma_0\rangle_\alpha \langle\sigma_y\rangle_\alpha\vert \le \epsilon
\end{equation}
where $\alpha$ is a pure state.  Essentially the same relation (using the same $\sigma_y$) holds for any spin $\sigma_x$ where~$\|x-y\|>R_\epsilon$.

%Because these inequalities holds for any $(\epsilon, R_\epsilon)$ pair, and because all two-spin correlations are equal in $\alpha$ and $\alpha'$, it follows that
%\begin{equation}
%\label{eq:cluster3}
%\langle\sigma_u\rangle_\alpha \langle\sigma_v\rangle_\alpha=\langle\sigma_u\rangle_{\alpha' }\langle\sigma_v\rangle_{\alpha'}\, 
%\end{equation}
%for any sites $u$ and $v$.

Squaring both sides of~(\ref{eq:cluster2}) and summing over $y$ gives 
\begin{equation}
\label{eq:cluster3}
\lim_{L\to\infty}\langle\sigma_0\rangle^2_\alpha\left(\frac{1}{\Lambda_L}\sum_{y\in\Lambda_L}\langle\sigma_y\rangle^2_\alpha\right)=\lim_{L\to\infty}\langle\sigma_0\rangle^2_{\alpha'}\left(\frac{1}{\Lambda_L}\sum_{y\in\Lambda_L}\langle\sigma_y\rangle^2_{\alpha'}\right)\, .
\end{equation}
In deriving~(\ref{eq:cluster3}) each sum was split into a finite sum over the spins within a range $R_\epsilon$ of the origin and an infinite sum over those outside. Because a sum over any finite set of spins cannot change the infinite sums in~(\ref{eq:cluster3}), for any choice of $\epsilon$ the equality in~(\ref{eq:cluster3}) holds to order~$\epsilon$. But because $\epsilon$ can be chosen to be any positive value, the equality in~(\ref{eq:cluster3}) follows. 

We then have
\begin{equation}
\label{eq:cluster4}
\langle\sigma_0\rangle^2_\alpha\ q^{(\alpha)}_{EA}= \langle\sigma_0\rangle^2_{\alpha'}\ q^{(\alpha')}_{EA}\, . 
\end{equation}
It has been proved~\cite{NRS23a} that if $\alpha$ and $\alpha'$ are chosen from the same mixed Gibbs state, then $q^{(\alpha)}_{EA}=q^{(\alpha')}_{EA}$. In the chaotic pairs picture, where $\alpha$ and $\alpha'$ must be chosen from different mixed states, equality of the EA order parameters must be added as an extra assumption.  In both cases this leads to
\begin{equation}
\label{eq:singlespin}
\langle\sigma_u\rangle_\alpha = \pm\langle\sigma_u\rangle_{\alpha'} 
\end{equation}
for any $u\in\mathbb{Z}^d$. 

Now suppose that $y$ is sufficiently far from $u$ and $v$ so that (with vanishingly small corrections), 
$\langle\sigma_u\rangle_\alpha\langle\sigma_y\rangle_\alpha=\langle\sigma_u\rangle_{\alpha'}\langle\sigma_y\rangle_{\alpha'}$ and 
$\langle\sigma_v\rangle_\alpha\langle\sigma_y\rangle_\alpha=\langle\sigma_v\rangle_{\alpha'}\langle\sigma_y\rangle_{\alpha'}$. If 
$\langle\sigma_y\rangle_\alpha=\langle\sigma_y\rangle_{\alpha'}$, then $\langle\sigma_u\rangle_\alpha=\langle\sigma_u\rangle_{\alpha'}$
and $\langle\sigma_v\rangle_\alpha=\langle\sigma_v\rangle_{\alpha'}$, while if $\langle\sigma_y\rangle_\alpha=-\langle\sigma_y\rangle_{\alpha'}$
then $\langle\sigma_u\rangle_\alpha=-\langle\sigma_u\rangle_{\alpha'}$ and $\langle\sigma_v\rangle_\alpha=-\langle\sigma_v\rangle_{\alpha'}$.
By appropriate choice of sites one can repeat this argument for all single-spin expectations to conclude that, if all two-spin correlations are equal in
$\alpha$ and $\alpha'$, then either $\langle\sigma_x\rangle_\alpha=\langle\sigma_x\rangle_{\alpha'}$ for all $x\in\mathbb{Z}^d$ or else
$\langle\sigma_x\rangle_\alpha=-\langle\sigma_x\rangle_{\alpha'}$ for all $x\in\mathbb{Z}^d$.

Once this has been established, higher-order odd correlations can be shown to be obey the same relations by taking derivatives with respect to couplings as before. This concludes the proof.~$\diamond$

\section{Nontrivial mixed Gibbs states and the metastate}
\label{sec:nontrivial}

In the following sections we will mostly (though not exclusively) be interested in PBC~metastates supported on Gibbs states which are mixtures of incongruent pure states.  In Sect.~\ref{sec:finite} we defined a nontrivial mixed state as a mixed thermodynamic state $\Gamma$ supported on more than one globally spin-reversed pair, i.e., $\Gamma=\sum_\alpha W_\alpha\alpha$ where $\alpha$ denotes a globally spin-reversed pure state pair, each individual $W_\alpha<1$  and $\sum_\alpha W_\alpha=1$. This decomposition is unique~\cite{Georgii11}. If the pure states form a continuum, a similar unique decomposition can be expressed as an integral (see Theorem 7.26 of~\cite{Georgii11}). Then we have the following result:

\medskip

{\bf Theorem 6.1 (Newman-Stein 2009)~\cite{NS09}.} Let $\kappa_J$ denote a PBC~metastate generated using the Hamiltonian~(\ref{eq:EA}) at fixed positive temperature. Then every Gibbs state~$\Gamma$ in the support of $\kappa_J$ must consist of either (a) a single pure state, or (b) a single pair of spin-reversed pure states, or (c) an infinity (which can be countable or uncountable) of incongruent pure state pairs.

\medskip

{\it Remark.\/} At zero temperature, $\kappa_J$ is supported on (one or more, depending on the scenario) trivial mixed states each consisting of a single ground state pair.

\medskip

We will also be interested in the metastate barycenter $\rho_J$ (also called the metastate averaged state~\cite{Read14}), defined as a single thermodynamic state $\rho_J=\int d\kappa_J(\Gamma) \Gamma$; i.e., the barycenter is an average over all thermodynamic states $\Gamma$ in the support of the metastate.  It is important to note that if the metastate is supported on more than a single Gibbs state~$\Gamma$, then in general the barycenter $\rho_J$ does not itself appear in the support of the metastate. The following result will also be relevant:

\medskip

{\bf Theorem 6.2 (Newman-Stein 2006)~\cite{NS06b}.} Suppose that the PBC metastate~$\kappa_J$ is supported on (one or more) nontrivial mixed state(s) $\Gamma$. Then the metastate barycenter $\rho_J$ is a single mixed Gibbs state supported on an uncountable infinity of pure states and whose decomposition has no atoms.

\medskip

%{\it Remark.\/} An immediate consequence of Theorem~6.2 is that if the nontrivial mixed states in $\kappa_J$ are each supported on a countable infinity of pure states, as in RSB, then $\kappa_J$ is supported on an uncountable infinity of mixed Gibbs states.

%\medskip

{\it Remark.} Read~\cite{Read14} has shown using field-theoretical techniques that the PBC~metastate constructed from RSB, each of whose Gibbs states are a mixture of a countable infinity of pure state pairs, has a barycenter $\rho_J$ which is a single mixed Gibbs state supported on an atomless continuum of pure state pairs, in accordance with Theorem~6.2.

%Of the four pictures for the low-temperature spin glass phase mentioned in the Introduction, three (droplet-scaling, TNT, and chaotic pairs) belong to class (b) of Theorem~6.1; only RSB belongs to class (c), i.e., its metastate is supported on nontrivial mixed Gibbs states. In the RSB picture each of these Gibbs states is predicted to comprise a countable infinity of pure state pairs, with each pair having nonzero weight; nonetheless the metastate barycenter $\rho_J$ is a single mixed Gibbs state supported on an atomless continuum of pure state pairs, indicating that in this picture $\kappa_J$ is supported on an uncountable infinity of distinct nontrivial mixed Gibbs states.

\section{Restricted metastate} 
\label{sec:restricted}

Our main interest in this paper is to study fluctuations in free energy differences between thermodynamic states that appear in the Edwards-Anderson PBC~metastate~$\kappa_J$.  Let $\Gamma$ and $\Gamma'$ be two infinite-volume Gibbs states, either pure or mixed, that appear in the support of $\kappa_J$ or its barycenter~$\rho_J$; any such Gibbs state is an element of ${\cal M}_1(\Sigma)$ that satisfies the DLR equations~\cite{Georgii11} for the Hamiltonian~(\ref{eq:EA}).   For these two infinite-volume states, we would like to understand the behavior of  $F_L(\Gamma,\Gamma')$, the difference in their free energies restricted to a finite volume~$\Lambda_L\subset\mathbb{Z}^d$:
\begin{equation}
\label{eq:diff}
\beta F_L(\Gamma,\Gamma')=\log\frac{\langle\exp(\beta H_\Lambda)\rangle_\Gamma}{\langle\exp(\beta H_\Lambda)\rangle_{\Gamma'}}\, ,
\end{equation}
where $H_\Lambda$ is the EA Hamiltonian restricted to $\Lambda_L$. The quantity $\langle\exp(\beta H_\Lambda)\rangle_\Gamma$ is a ratio whose numerator contains a Boltzmann factor on
spin configurations only in $\mathbb{Z}^d\backslash\Lambda$, the complement of $\Lambda$ in $\mathbb{Z}^d$, while the denominator is the Boltzmann factor on all infinite-volume spin configurations. (While the numerator and denominator of $\langle\exp(\beta H_\Lambda)\rangle_\Gamma$ are individually not well-defined, the ratio is well-defined.)
$F_L(\Gamma,\Gamma')$ represents the difference in free energies within $\Lambda_L$ between boundary conditions chosen from the infinite-volume Gibbs states $\Gamma$ and $\Gamma'$, respectively~\cite{ANSW14}.

In order to study this quantity, we construct a new type of metastate that groups together Gibbs states having a common overlap with a reference pure state; we will refer to this new type of metastate as a {\it restricted metastate\/}. To begin we choose a pure state $\omega$ randomly from~$\kappa_J$ as follows: first, choose a Gibbs state~$\Gamma$ from the distribution $\kappa_J(\Gamma)$. If $\Gamma$ is itself pure or else an equal mixture of a spin-reversed pair of pure states, then $\omega=\Gamma$ in the first case or $\omega$ equals either of the two pure states in the second. (Because our focus will be on edge overlaps, which of the two pure states is chosen will be immaterial.) On the other hand, if $\Gamma$ is a nontrivial mixture of infinitely many pure states, then one chooses a pure state $\omega$ from $\Gamma$ according to its pure state decomposition. If the decomposition into pure states is not countable, then one chooses a pure state according to the integral decomposition~(7.26) in~\cite{Georgii11} (we remark that the theorem in~\cite{Georgii11} is a special case of the more general Choquet theory~\cite{Phelps66}). This procedure is equivalent to choosing a pure state randomly from the metastate barycenter~\cite{NS06b}.  We also choose an interval $(p-\delta,p+\delta)$ with $p\in(-1,1)$, $\delta>0$ and 
\begin{equation}
\label{eq:delta}
\delta\ll\begin{cases}{\rm min}(p,1-p) & p>0\, ,\\
{\rm min}(1+p,-p)&p<0\, ,\\
1& p=0\, .{\normalsize }
\end{cases}
\end{equation}

We now present two constructions which may give rise to different restricted metastates, but which will both satisfy the three properties listed in Sect.~\ref{sec:meta}.

{\it Construction~1\/.}  In this construction the restricted metastate consists of all pure states in $\kappa_J$ whose edge overlap $q^{(e)}_{\alpha\omega}$ with $\omega$ is within the predetermined restricted interval introduced above. To accomplish this, for every Gibbs state $\Gamma$ in $\kappa_J$ one retains only those pure states $\alpha$ in the support of $\Gamma$ satisfying $q^{(e)}_{\alpha\omega}\in(p-\delta,p+\delta)$; the remaining pure states are discarded. (If no pure state in $\Gamma$ satisfies this condition, $\Gamma$ itself is discarded.) The weights of the remaining $\alpha$'s are then rescaled so that the rescaled weights add up to one with their relative probabilities the same as before. This procedure is carried out for every Gibbs state $\Gamma$ in the support of $\kappa_J$. When this is done we renormalize the overall mass to compensate for the discarded~$\Gamma$'s. 

Construction~1 may be used if the overlap distribution of $\rho_J$, the barycenter of $\kappa_J$, is spread over a nonzero interval. However, an important special case, which is relevant to RSB (see, e.g.~\cite{Parisi96,Read14}), is that the overlap distribution of $\rho_J$ is a single $\delta$-function, in which case Construction~1 may (depending on $p,\delta$) discard a set of $\Gamma$'s with measure one in the metastate.  To avoid this situation, one would then use Construction~2.

{\it Construction 2.\/}  For a given $\omega$ one retains only the $\Gamma$ from which $\omega$ was chosen~\cite{omeganote}, and then follows the procedure of Construction~1 for each of the pure states in $\Gamma$.

The discussion so far outlines a procedure that uses a fixed $\omega$ chosen from the PBC metastate. In order to construct a new metastate, {\it every\/} pure state $\omega$ in~$\rho_J$ needs to be considered as a possible reference pure state.  Consequently, $\omega$ itself is treated as a random variable chosen from~$\rho_J$. The resulting object is a~$(p,\delta)$-restricted measure ${\kappa}^{p,\delta}_{J,\omega}$ on Gibbs states; the notation is chosen to separate $p$ and $\delta$, which are fixed parameters, from $J$ and $\omega$, which are random quantities.   For either of the two constructions described above, if the barycenter~$\rho_J$ of the original PBC metastate~$\kappa_J$ consists of an uncountable mixture of pure states, then the restricted metastate constructed from $\kappa_J$ will also be supported on an uncountable set of mixed states, albeit in a novel way, as $\omega$ varies.

We then have the following theorem:

\medskip

{\bf Theorem 7.1.} At any positive temperature, ${\kappa}^{p,\delta}_{J,\omega}$ as constructed above satisfies the three conditions for a translation-covariant metastate, but now depending on both $J$ and $\omega$.  

\medskip

{\bf Proof.}  The first condition to be satisfied is that the restricted metastate ${\kappa}^{p,\delta}_{J,\omega}$ be a probability measure on Gibbs states. By the method of construction, ${\kappa}^{p,\delta}_{J,\omega}$ is supported solely on mixtures of pure states appearing in the support of the PBC metastate, and the renormalization of its mass guarantees that the probabilities of all states add up to one.  It remains to show that the properties of coupling-covariance and translation-covariance are also satisfied by ${\kappa}^{p,\delta}_{J,\omega}$.

Lemma~4.1 ensures that coupling covariance, which now takes the form
\begin{equation}
\label{eq:cc}
{\kappa}^{p,\delta}_{J+J_B,{\cal L}_{J_B}\omega}(A)= {\kappa}^{p,\delta}_{J,\omega}({\cal L}_{J_B}^{-1}A)\, ,
\end{equation}
is satisfied.  

Translation-covariance follows because $\omega$ is treated as a random variable rather than as a fixed state; i.e., an event, which in this setting is a function on the spins, is evaluated in terms of its $(J,\omega)$-probability.  Eq.~(\ref{eq:3}) is then replaced by
\begin{equation}
\label{eq:modcov}
{\kappa}^{p,\delta}_{\tau J,\tau\omega}(A)={\kappa}^{p,\delta}_{J,\omega}(\tau^{-1}A)
\end{equation}
which, given the translation covariance of Gibbs states in $\kappa_J$, is clearly satisfied. $\diamond$

\medskip
%
%\begin{itemize}
%\item The pure states from the support of ${\tilde\kappa}^{p,\delta}_{J,\omega}$ are a subset of those from the support of a single PBC metastate~$\kappa_J$.
%\item The pure state pairs from the support of ${\tilde\kappa}^{p,\delta}_{J,\omega}$ are mutually incongruent; this follows from a theorem in~\cite{NS01c}.
%\item The edge overlap of any pure state from the support of ${\tilde\kappa}^{p,\delta}_{J,\omega}$ with an external reference state $\omega$ lies in the interval $(p-\delta,p+\delta)$.
%\end{itemize}
%The restricted metastate is primarily useful in studying systems whose thermodynamic structure comprises nontrivial mixed Gibbs states. 
 In droplet-scaling and TNT, the PBC metastate~$\kappa_J$ is supported on a single pair of spin-reversed pure states. In these cases, either a restricted metastate is not defined (if $q^{(e)}_{EA}\notin[p-\delta,p+\delta]$) or else it simply maps $\kappa_J$ back to itself.  On the other hand, if the support of the metastate includes incongruent pure states, as in RSB and chaotic pairs, then the restricted metastate is well-defined for (some or all) values of $p\ne q^{(e)}_{EA}$. (In the case of chaotic pairs, Construction~1 must be used, or else the restricted metastate will again be trivial.)

\section{Metastate incongruence}
\label{sec:metainc}

From here on we write $(J,\omega)$ for a random pair consisting of a coupling realization and a (random) choice of $\omega$ for that individual~$J$.  As before, we assume a PBC metastate~$\kappa_J$ supported on incongruent pure state pairs. 

\medskip

{\bf Lemma 8.1.} Given a PBC metastate $\kappa_J$, for each~$(J,\omega)$ pair the edge overlap $\overline{\langle\sigma_x\sigma_y\rangle_\Gamma\langle\sigma_x\sigma_y\rangle_{\omega}}=p+O(\delta)$ for any Gibbs state~$\Gamma$ in the support of the restricted metastate~${\kappa}^{p,\delta}_{J,\omega}$.

\medskip

{\bf Proof.} To show this it is convenient to use the notation ${\tilde\kappa}^{p,\delta}_{J,{\hat\omega}}$ when the $\omega$ in ${\kappa}^{p,\delta}_{J,\omega}$ is no longer random but equals a particular $\hat\omega$. Consider a Gibbs state~$\Gamma=\sum_\alpha W_\alpha \alpha$ in the support of~${\tilde\kappa}^{p,\delta}_{J,\hat\omega}$; the $W_\alpha$'s correspond to the weights of the pure states in the decomposition of $\Gamma$, and add up to one. By construction, every pure state $\alpha\in\Gamma$ has  $\overline{\langle\sigma_x\sigma_y\rangle_\alpha\langle\sigma_x\sigma_y\rangle_{\hat\omega}}=p+r_\alpha$ where $\vert r_\alpha\vert\le\delta$. Therefore
\begin{equation}
\label{eq:Gamma1}
\overline{\langle\sigma_x\sigma_y\rangle_\Gamma\langle\sigma_x\sigma_y\rangle_{\hat\omega}}=\sum_\alpha W_\alpha\overline{\langle\sigma_x\sigma_y\rangle_\alpha\langle\sigma_x\sigma_y\rangle_{\hat\omega}}=p\sum_\alpha W_\alpha+\sum_\alpha W_\alpha r_\alpha=p+\sum_\alpha W_\alpha r_\alpha
\end{equation}
 so that
\begin{equation}
\label{eq:Gamma2}
p-\delta\le\overline{\langle\sigma_x\sigma_y\rangle_\Gamma\langle\sigma_x\sigma_y\rangle_{\hat\omega}}\le p+\delta
\end{equation}
for any~$\Gamma\in{\tilde\kappa}^{p,\delta}_{J,\hat\omega}$. This argument holds for each~choice of $\hat\omega$, and the conclusion follows. $\diamond$

\medskip

Next consider the average of the overlap $\overline{\langle\sigma_x\sigma_y\rangle_\Gamma\langle\sigma_x\sigma_y\rangle_{\hat\omega}}$ for $\Gamma$ in~${\tilde\kappa}^{p,\delta}_{J,{\hat\omega}}$. Using~(\ref{eq:Gamma2}) we have
\begin{eqnarray}
\label{eq:spaceavg}
{\tilde\kappa}^{p,\delta}_{J,\hat\omega}\Bigl(\overline{\langle\sigma_x\sigma_y\rangle_\Gamma\langle\sigma_x\sigma_y\rangle_{\hat\omega}}\Bigr)&:=&\int d{\tilde\kappa}^{p,\delta}_{J,\hat\omega}(\Gamma)\overline{\langle\sigma_x\sigma_y\rangle_\Gamma\langle\sigma_x\sigma_y\rangle_{\hat\omega}}\nonumber\\
&=&p+O(\delta)\, .
\end{eqnarray}

\medskip

Consider now a joint construction of two restricted metastates generated in the following manner: for each reference pure state~$\hat\omega$ chosen from the barycenter~$\rho_J$, construct two metastates (using either Construction~1 or~2 for each as appropriate) with two different values of $p$; call them $p_1$ and $p_2$.  The  resulting ${\tilde\kappa}^{p_1,\delta}_{J,\hat\omega}$ and ${\tilde\kappa}^{p_2,\delta}_{J,\hat\omega}$ have $0\le p_1<p_2<1$ and $0<\delta< {\rm min}(p_1,p_2-p_1, 1-p_2)$.  Using~(\ref{eq:spaceavg}) we then have 
 \begin{eqnarray}
 \label{eq:r1}
{\tilde\kappa}^{p_1,\delta}_{J,\hat\omega}\Bigl(\overline{\langle\sigma_x\sigma_y\rangle_\Gamma\langle\sigma_x\sigma_y\rangle_{\hat\omega}}\Bigr)=p_1 + O(\delta)\nonumber\\
\ne{\tilde\kappa}^{p_2,\delta}_{J,\hat\omega}\Bigl(\overline{\langle\sigma_x\sigma_y\rangle_\Gamma\langle\sigma_x\sigma_y\rangle_{\hat\omega}}\Bigr)=p_2 + O(\delta)\, .
\end{eqnarray}

The inequality~(\ref{eq:r1}) can hold only if ${\tilde\kappa}^{p_1,\delta}_{J,\hat\omega}\Bigl(\langle\sigma_x\sigma_y\rangle_\Gamma\langle\sigma_x\sigma_y\rangle_{\hat\omega}\Bigr)\ne{\tilde\kappa}^{p_2,\delta}_{J,\hat\omega}\Bigl(\langle\sigma_x\sigma_y\rangle_\Gamma\langle\sigma_x\sigma_y\rangle_{\hat\omega}\Bigr)$ for a positive density of edges. Because the choice of $\hat\omega$ is the same for both metastates, it must then also be true that ${\tilde\kappa}^{p_1,\delta}_{J,\hat\omega}(\langle\sigma_x\sigma_y\rangle_\Gamma)\ne{\tilde\kappa}^{p_2,\delta}_{J,\hat\omega}(\langle\sigma_x\sigma_y\rangle_\Gamma)$ for a positive density of edges. Because this is so for each~instance of $(J,\hat\omega)$, it follows that for any edge~$(x,y)$
\begin{equation}
\label{eq:incong2}
(\nu\times\kappa_J)\Bigl\{(J,\omega):{\kappa}^{p_1,\delta}_{J,\omega}\big(\langle\sigma_x\sigma_y\rangle_\Gamma\big)\neq {\kappa}^{p_2,\delta}_{J,\omega}\big(\langle\sigma_x\sigma_y\rangle_\Gamma\big)\Bigr\}>0\, ,
\end{equation}
where $\nu\times\kappa_J$ denotes $\nu(dJ)\kappa_J(d\omega)$.

 We may now apply Theorem~4.2 of~\cite{ANSW14}, which in the present context can be expressed as: 

\medskip

{\bf Theorem 8.2} (modified from~\cite{ANSW14}):  Consider two infinite-volume (pure or mixed) Gibbs states $\Gamma$ and $\Gamma'$ chosen from distinct restricted metastates satisfying~(\ref{eq:incong2}), and let~$F_L(\Gamma,\Gamma')$ denote their free energy difference restricted within a volume~$\Lambda_L=[-L,L]^d\subset\mathbb{Z}^d$ (for a formal definition, see Eq.~(3) in~\cite{ANSW14}) . Then 
there is a constant $c>0$ such that the variance of $F_L(\Gamma,\Gamma')$ under the probability measure $M:=\nu(dJ)\kappa_J(d\omega){\kappa}^{p_1,\delta}_{J,\omega}(d\Gamma)\times{\kappa}^{p_2,\delta}_{J,\omega}(d\Gamma')$ satisfies
\begin{equation}
\label{eq:flucs}
{\rm Var}_M\Big(F_L(\Gamma,\Gamma')\Big)\ge c\vert\Lambda_L\vert\, .
\end{equation}

\section{Free energy difference fluctuations between pure states}
\label{sec:pure}

In this section we use the restricted metastate to explore fluctuations in the free energy difference between pure states chosen from a PBC metastate barycenter~$\rho_J$. There are two cases to consider:
free energy differences between pure states chosen from the {\it same\/} mixed state $\Gamma$ in the support of the PBC metastate $\kappa_J$, and those between pure states chosen from {\it different\/} mixed states $\Gamma$ and $\Gamma'$ in the support of $\kappa_J$. We begin with the former.

\subsection{Pure states chosen from the same mixed Gibbs state}
\label{subsec:same}

We consider first fluctuations in the free energy difference between two pure states within the same mixed~$\Gamma$. Suppose that in some~$\Gamma$ there are at least two distinct non-self overlap values, $p_1$ and $p_2$. Then one can always find three incongruent pure states $\alpha$, $\beta$, and $\omega$ in that~$\Gamma$ for which $q^{(e)}_{\alpha\omega}=p_1$ and $q^{(e)}_{\beta\omega}=p_2$. (Indeed, all three will be in the same $\Gamma$ as noted earlier if Construction~2 is used.) Thus $\alpha$ and $\beta$ would belong to different restricted metastates as in~(\ref{eq:incong2}), and the lower bound~(\ref{eq:flucs}) applies.

It could also be the case that, as in 1-step RSB, there is only a single non-self-overlap value $q^{(e)}_0<q_{EA}^{(e)}$. In that case one chooses $p_1=q_0^{(e)}$ and $p_2=q_{EA}^{(e)}$. Using Construction~2 with these values for the two restricted metastates, one can then apply the lower bound~(\ref{eq:flucs}).

There is also an upper bound following from a known result on the free energy itself~\cite{WA90}, which when applied to the situation considered here can be stated as
\begin{equation}
\label{eq:flucsupper}
{\rm Var}_M\Big(F_L(\Gamma,\Gamma')\Big)\le d\vert\Lambda_L\vert\, ,
\end{equation}
where $d>0$ is again positive and independent of the volume. We therefore have

{\bf Theorem 9.1}. For any two incongruent pure states $\alpha$ and $\alpha'$ appearing in the pure state decomposition of a single nontrivial mixed state~$\Gamma$ in the support of $\kappa_J$, there exist constants $0<c\le d$ such that
\begin{equation}
\label{eq:bounds}
c\vert\Lambda_L\vert\le{\rm Var}_M\Big(F_L(\alpha,\alpha')\Big)\le d\vert\Lambda_L\vert\, .
\end{equation}

\subsection{Pure states chosen from different mixed Gibbs states}
\label{subsec:different}

Here we must consider two possible cases. The first corresponds to situations where the PBC metastate barycenter $\rho_J$ has a nontrivial edge overlap distribution spread over a range of multiple values. The second, relevant to the RSB scenario, is where the pure state overlap distribution of~$\rho_J$ is concentrated on a single value and is therefore a delta-function at that value. 

In the first case we can use Construction~1 from Sect.~\ref{sec:restricted}. Here the argument goes through exactly as for two pure states from the same mixed Gibbs state and the conclusion, i.e., Theorem 8.2, applies. The second case requires additional discussion.

Suppose that the barycenter edge overlap distribution $P_{\rho_J}(q)=\delta(q-q_0)$ with $q_0<q^{(e)}_{EA}$, where $q^{(e)}_{EA}$ is the edge self-overlap (assumed to be the same for all mixed Gibbs states in the support of $\kappa_J$); the inequality $q_0<q^{(e)}_{EA}$ follows from Cauchy-Schwartz.  A reference pure state~$\omega$ is chosen in the usual way (which is the same for either construction). One then uses Construction~1 with $p=q_0$ to build the first metastate~$\kappa^{q_0,\delta}_{J,\omega}$ and Construction~2 to build the second metastate~$\kappa^{p,\delta}_{J,\omega}$, where $p\notin[q_0-\delta,q_0+\delta]$. Then with probability~one (in $({\cal J},\omega)$) the two constructions will contain pure states in different mixed Gibbs states, and the argument goes through as before.

Up until now the discussion has focused on metastates supported on nontrivial mixtures of multiple pure state pairs, but here we may also draw conclusions about scenarios, such as chaotic pairs, in which $\kappa_J$ is supported on multiple Gibbs states each supported on a trivial mixture of a pair of spin-reversed pure states. Pure states chosen from different trivially mixed Gibbs states must also be incongruent as a result of Theorem~5.2. In this case one uses Construction~1, with $p_1=q^{(e)}_{EA}$ and $p_2$ set equal to the edge overlap value(s) of $\omega$ with pure states chosen from other Gibbs states, to construct two incongruent restricted metastates, to which Theorem~8.2 may then be applied.

This demonstrates that the conclusion of Theorem~9.1 is not limited to PBC~metastates supported on Gibbs states which are nontrivial mixtures of many pure state pairs: it applies to any $\kappa_J$ whose support contains incongruent pure states regardless of how they're arranged in mixed Gibbs states.  We can combine the results of this and the previous section in the following theorem:

\medskip

{\bf Theorem 9.2.} Suppose that in the EA Ising model there is a phase with multiple incongruent pure state pairs. Then for any incongruent pure states $\alpha$ and $\alpha'$ chosen from the metastate barycenter~$\rho_J$, the variance ${\rm Var}_M\Big(F_L(\alpha,\alpha')\Big)$ of their free energy difference satisfies~(\ref{eq:bounds}).

\section{Free energy difference fluctuations between mixed Gibbs states}
\label{sec:mixed}

We now turn to free energy differences between entire mixed Gibbs states which can be decomposed into pure states according to~$\Gamma=\sum_\alpha W_\alpha \alpha$, where, as discussed in Sect.~\ref{sec:metainc}, the weights $W_\alpha$ satisfy $\sum_\alpha W_\alpha =1$.  In considering edge overlaps between mixed Gibbs states, an issue arises: although by Lemma~4.1 edge overlaps between distinct pure states are invariant under finite changes of couplings, the individual pure state weights are not~\cite{NS03b}; under a change in coupling $J_{xy}\to J_{xy}+\Delta J$, the weight of pure state $\alpha$ changes according to~\cite{AW90,NSBerlin,NS97,NS98,NS02,NS03b}
\begin{equation}
\label{eq:weights}
W_\alpha\to W'_\alpha=r_\alpha W_\alpha/\sum_\gamma r_\gamma W_\gamma\, ,
\end{equation}
where
\begin{equation}
\label{eq:ralpha}
r_\alpha = \Bigl\langle\exp(\beta\Delta J\sigma_x\sigma_y\Bigr\rangle_\alpha\, .
\end{equation}

In the general case where $\kappa_J$ contains incongruent states, the edge overlap $q^{(e)}_{\omega\Gamma}=\sum_{\alpha\in\Gamma} W_\alpha q^{(e)}_{\omega\alpha}$ between a pure state $\omega$ and a mixed Gibbs state $\Gamma$ that does not contain $\omega$ transforms when $J_{xy}\to J_{xy}+\Delta J$ as
\begin{equation}
\label{eq:pureoverlap}
q^{(e)}_{\omega'\Gamma'}=\frac{1}{\sum_\gamma r_\gamma W_\gamma} \sum_{\alpha\in\Gamma}r_\alpha W_\alpha\ q^{(e)}_{\omega\alpha}\ne q^{(e)}_{\omega\Gamma}
\end{equation}
in general if $q^{(e)}_{\omega\alpha}$ varies with $\alpha$.  

In the chaotic pairs picture, where each Gibbs state is a mixture of two spin-reversed pure states each with weight 1/2, the weights do not change under the transformation~(\ref{eq:pureoverlap}), so edge overlaps between entire mixed states remain invariant under a finite change in couplings and Theorem~9.2 again applies, this time to mixed states. However, when nontrivial mixed Gibbs states appear as in RSB, then under a finite coupling change the edge overlap will in general change as a consequence of~(\ref{eq:pureoverlap}). Consequently, in this scenario restricted metastates cannot be constructed using overlaps between a pure state and a mixed state (or between two mixed states) because coupling covariance is lost.

Nevertheless, as discussed earlier an important special case is that where the edge overlap distribution of~$\rho_J$ consists of a single $\delta$-function, as is predicted to occur in the RSB~scenario~\cite{Parisi96,Read14}.  Suppose this is the case; as before, we denote this single edge overlap value as~$q_0$, and as before $\vert q_0\vert<q^{(e)}_{EA}$. Then consider two distinct mixed Gibbs states  $\Gamma=\sum_\alpha W_\alpha \alpha$ and $\Gamma'=\sum_{\alpha'} W_{\alpha'}\alpha'$.  A natural question (which is relevant only in situations where the $\Gamma$'s are nontrivial mixtures of pure state pairs) arises: can the decompositions of $\Gamma$ and $\Gamma'$ include any pure states common to the two?

We know of no argument that forbids this in the general case. However, in a scenario where the mixed Gibbs states are each supported on a countable infinity of pure states, $\kappa_J$ is supported on an uncountable infinity of mixed states (see the remark following Theorem~6.2). In this case the answer to the above question is no, in the sense that a single pure state cannot be shared by a set of $\Gamma$'s with positive measure in~$\kappa_J$. For if this were so, that pure state (which has positive weight in each of the $\Gamma$'s in which it appears) would then have positive weight in $\rho_J$, which violates Theorem~6.2. 

%and the overlap distribution of the barycenter $P_{\rho_J}=\delta(q-q_0)$ (with, as before $q_0<q^{(e)}_{EA}$) 

This argument doesn't rule out a given pure state from appearing in a set of mixed states with {\it zero\/} measure in $\kappa_J$. But because the union of a countable collection of sets having zero measure itself has zero measure~\cite{Royden10}, it does imply that a given mixed state can share any or all of its pure states with a set of other mixed states having at most zero measure in~$\kappa_J$. The conclusion is that if one chooses a $\Gamma$ and $\Gamma'$ independently at random from $\kappa_J$, then with probability~one they have no pure states in common.  If $\Gamma$ and $\Gamma'$ have no pure states in common and the overlap distribution $P_{\rho_J}(q)=\delta(q-q_0)$, then their edge overlap is
\begin{equation}
\label{eq:mixedoverlap}
q^{(e)}_{\Gamma\Gamma'}=\sum_{\alpha\in\Gamma,\alpha'\in\Gamma'}W_\alpha W_{\alpha'}q^{(e)}_{\alpha\alpha'}=\sum_{\alpha\in\Gamma,\alpha'\in\Gamma'}W_\alpha W_{\alpha'}q_0=q_0\, .
\end{equation}
In this case $q^{(e)}_{\Gamma\Gamma'}$ is again invariant under changes in couplings, and restricted metastates can now be constructed.

There are two possible ways of doing this, which lead to the same result. The first is to proceed as before by choosing a pure state~$\omega$ randomly from $\rho_J$; the overlap~$q^{(e)}_{\omega\Gamma'}$ with a mixed state $\Gamma'$ that does not contain~$\omega$ will again be $q_0$ by a computation similar to that in~(\ref{eq:mixedoverlap}). The second is to begin by choosing a mixed state~$\Gamma$ from the support of $\kappa_J$, and use that as the reference state.  For specificity we will choose the second method, but it will become clear that which is chosen makes no difference.

Another distinction with the procedures used in~Sect.~\ref{sec:restricted} is that here we extend the range of possible values of the parameter~$\delta$ to include zero (see the discussion around Eq.~(\ref{eq:delta})). We then construct two metastates as follows: for a given reference Gibbs state~$\Gamma$, use Construction~1 with $p=q_0$, $\delta=0$. The resulting ${\kappa}^{q_0,0}_{J,\Gamma}$ will include a set of Gibbs states having measure~one in~$\kappa_J$. It will not include the reference state~$\Gamma$, because the $\kappa_J$-probability that $q^{(e)}_{\Gamma\Gamma}=q_0$ is zero. To see that this is so, note that if $q^{(e)}_{\Gamma\Gamma}=q_0$ for some $J$, an infinitesimal change in a single coupling will shift all the pure state weights in $\Gamma$, thereby changing $q^{(e)}_{\Gamma\Gamma}$ while leaving the overlap distribution of $\rho_J$ unchanged~\cite{NS96b}.

For the second metastate, one uses Construction~2 with $p$ lying in either of the two intervals $[-1,q_0)$ or $(q_0,1]$.  At this point the rest of the argument in~Sect.~\ref{sec:metainc} goes through exactly as before; as a consequence, Theorem 9.1 then applies (with probability one) to nontrivial mixed Gibbs states as well.

Free energy differences between mixed Gibbs states are particularly important in consideration of spin glass stiffness, which is related to free energy differences in finite systems under a change in boundary conditions, such as periodic to antiperiodic.  This amounts to a change in the mixed Gibbs state seen inside a window (in the sense of correlations) far from the boundaries. We will examine this further in a separate study~\cite{MNSinprep}.

\section{Energy difference fluctuations at zero temperature}
\label{sec:zero}

Up until now the discussion has been confined exclusively to free energy difference fluctuations at positive temperature. Aside from the natural question of whether and to what extent these results
extend to zero temperature, consideration of zero temperature behavior may be useful for addressing several issues. One is whether the zero-temperature PBC~metastate is supported on a single ground state pair (as a function of dimension) or multiple incongruent ground state pairs, and another concerns comparison to numerical studies of spin glass stiffness which are often (though not exclusively) performed at zero temperature.

A straightforward construction of restricted metastates at zero temperature is not feasible in the general case, for a reason similar (but with a different cause) to that encountered in the previous section. The statement and proof of Lemma~4.1 are limited to strictly positive temperature. At zero temperature a change in a single coupling could --- in principle --- lead to a new ground state related to the original by a droplet flip whose boundary contains a positive density set of couplings (including of course the altered coupling). Whether such infinite positive-density droplets exist is unknown, but they are expected in any dimension where RSB describes the low-temperature phase~\cite{NS22}.  If such droplets do exist, and one changes a coupling leading to a positive-density droplet flip in one GSP but not another, their edge overlap could change. Consequently the issue of coupling covariance again becomes relevant.

However, there is an important special case where restricted metastates can be constructed at zero temperature.  We first note that, regardless of the specific form of the low-temperature phase, the PBC metastate~$\kappa_J$ of the EA~Hamiltonian~(\ref{eq:EA}) at zero temperature is a probability measure on mixed Gibbs states each consisting of a single spin-reversed ground state pair (GSP), with a GSP denoted~$\alpha$. In droplet-scaling and TNT, $\kappa_J$ at zero temperature is supported on a single~GSP while in RSB and chaotic pairs it's supported on multiple incongruent~GSP's.

Supposing that $\kappa_J$ is supported on multiple GSP's, one can proceed if the edge overlap distribution~$P_{\rho_J}(q)$ of the barycenter~$\rho_J$ consists of a single $\delta$-function, i.e.,  $P_\mu(q)=\delta(q-q_0)$ where now $\vert q_0\vert<1$. As before, $q_0$ is constant for a.e.~$J$~\cite{NS96b}. In this situation any two incongruent ground states chosen randomly from the PBC metastate will have $\nu(dJ)\kappa_J(d\alpha)$-probability one of having overlap~$q_0$.  Consequently the edge overlap remains invariant with probability~one upon a finite change of couplings, and coupling covariance is restored.

The procedure for constructing two incongruent restricted metastates is similar to that used in Sect.~\ref{sec:mixed}. Given a reference GSP~$\omega$ chosen randomly from $\kappa_J$, for the first metastate use Construction~1 with $p=q_0$, $\delta=0$. If $\kappa_J$ is supported on a countable set of GSP's, the resulting ${\kappa}^{q_0,0}_{J,\omega}$ will include all GSP's in the support of~$\kappa_J$ except $\omega$;  if $\kappa_J$ is supported on an uncountable set of GSP's, then ${\kappa}^{q_0,0}_{J,\omega}$ will include a set of GSP's having measure~one in~$\kappa_J$ but excluding~$\omega$. The second metastate is then constructed using Construction~2 with $p=1$ and $\delta=0$, and will contain only~$\omega$. Then an argument similar to that leading to~(\ref{eq:incong2}) can be used to show the two resulting metastates are incongruent, and the bound~(\ref{eq:flucs}) again holds (with probability one in $\nu(dJ)\kappa_J(d\alpha)$) for incongruent ground states.

\section{Gaussian behavior of the free energy difference distribution}
\label{sec:Gaussian}

In previous sections it was shown that the variance of the free energy difference within a restricted volume~$\Lambda_L\subset{\mathbb Z}^d$ of two pure or mixed Gibbs states chosen from the PBC~metastate $\kappa_J$ (or its barycenter~$\rho_J$) scales as~$\vert\Lambda_L\vert$; roughly speaking, their free energy difference within a volume of size $L^d$ scales as $L^{d/2}$. The natural next step is to ask whether one can say more about the distribution of this difference.   This is possible by using a partially averaged version of the free energy difference.  Let $M(F_L(\Gamma,\Gamma')\vert J_L)$ denote the expectation under $M$ (defined in Theorem~8.2) of the free energy difference between $\Gamma$ and $\Gamma'$ conditioned on the couplings $J_L$ inside $\Lambda_L$.  Then it was proved in~\cite{ANSW14} (using a result from~\cite{AW90}) that the distribution of $M(F_L(\Gamma,\Gamma')\vert J_L)/\sqrt{\vert\Lambda_L\vert}$ has as $L\to\infty$ (at least) a Gaussian tail; that is, there exists a $c>0$ such that for all~$t$,
\begin{equation}
\label{eq:root}
\liminf_{L\to\infty}\nu\Bigl(\exp t\ \frac{M(F_L(\Gamma,\Gamma')\vert J_L)}{\sqrt{\vert\Lambda_L\vert}}\Bigr)\ge e^{ct^2}\, ,
\end{equation}
in any dimension in which $\Gamma$ and $\Gamma'$ are chosen from incongruent metastates. 

These results lead to new insights in two dimensions, where extensive numerical studies strongly indicate that $\kappa_J$ is supported on a single (paramagnetic) pure state at all positive temperature~\cite{Rieger96,HY01,Houdayer01,CBM02}.  In two dimensions $\sqrt{\vert\Lambda_L\vert}$ and $\vert\partial\Lambda_L\vert$ have the same scaling with $L$, so~(\ref{eq:root}) can be replaced by
\begin{equation}
\label{eq:lower}
\liminf_{L\to\infty}\nu\Bigl(\exp t\ \frac{M(F_L(\Gamma,\Gamma')\vert J_L)}{\vert\partial\Lambda_L\vert}\Bigr)\ge e^{ct^2}\, .
\end{equation}
It was further proved in~\cite{ANSW14} that, for $\beta<\infty$
\begin{equation}
\label{eq:upper}
\limsup_{L\to\infty}\nu\Bigl(\exp t\ \frac{M(F_L(\Gamma,\Gamma')\vert J_L)}{\vert\partial\Lambda_L\vert}\Bigr)\le e^{4\beta t}\, ,
\end{equation}
which follows from the fact that the free energy difference of two Gibbs states in any sequence of volumes $\Lambda_L$ cannot scale faster than their surface area $\vert\partial\Lambda_L\vert$. %Analogies to finite systems suggest the possibility of a stronger upper bound~\cite{AFunpub,NSunpub,Stein16}, but none have been found as yet.

An immediate consequence of~(\ref{eq:lower}) and~(\ref{eq:upper}), which are in contradiction for sufficiently large~$t$, combined with the results of Sect.~\ref{sec:pure}, is as follows: on ${\mathbb Z}^2$, for any fixed $\beta<\infty$, the support of~$\kappa_J$ cannot include incongruent Gibbs states; at most the support of $\kappa_J$ can be a single spin-reversed pure state pair.  To the best of our knowledge this is a first (though modest) step toward a full proof of a unique Gibbs state at all positive temperature in two dimensions.

We can also gain some insight into zero-temperature behavior in two dimensions, where the question of whether $\kappa_J$ is supported on a single GSP or multiple (necessarily incongruent) GSP's remains an open question~\cite{NS2D00,NS2D01,ADNS10,Fan23}. There exists a proof that $\kappa_J$ for the EA model at zero temperature on the {\it half\/}-plane $y\ge 0$, with free boundary conditions on $y=0$, is supported on a single~GSP~\cite{ADNS10}, but this has not yet been extended to the full plane.   Using the results above (noting that when $\beta=\infty$ the RHS of~(\ref{eq:upper}) is replaced by $e^{4t}$) we have the following partial result: at zero temperature in two dimensions, the support of $\kappa_J$ cannot include incongruent GSP's with an overlap distribution~$P_{\rho_J}(q-q_0)$.
%$E\Bigl[F_L(\Gamma,\Gamma')\Big|J_L\Bigr]-E\Bigl[F_L(\Gamma,\Gamma')\Bigr]$, where $E[\cdot]$ is an average over all couplings and $E[\cdot\vert J_L]$ is the conditional expectation with the conditioning done on the couplings within the volume $\Lambda_L$. Clearly ~$E\bigl[\tilde{F}_L(\Gamma,\Gamma')\bigr]=0$; $\tilde{F}_L(\Gamma,\Gamma')$ is in some rough sense a metastate average over the free energy difference within a volume~$\Lambda_L$.  
%Then it follows from Proposition~6.1 of~\cite{AW90} that the distribution of $\tilde{F}_L(\Gamma,\Gamma')$ is a Gaussian with mean zero and variance equal to~$b\vert\Lambda_L\vert$, where $b>0$ is independent of $L$.

\section{Summary and Discussion}
\label{sec:discussion}

In this paper we have introduced a framework for studying fluctuations of the free energy difference between two infinite-volume Gibbs states of the Edwards-Anderson Hamiltonian in arbitrary finite dimension.  We introduce a new type of metastate, called the {\it restricted metastate\/}, which classifies (pure or mixed) Gibbs states according to their edge overlaps with a reference (pure or mixed) Gibbs state; all of these states are randomly drawn from a periodic boundary condition metastate $\kappa_J$ constructed from the EA Hamiltonian~(\ref{eq:EA}) at fixed inverse temperature $\beta$.

Using the restricted metastate construction(s) one can choose two incongruent pure states appearing in the support of $\kappa_J$ (and therefore in its barycenter~$\rho_J$), and place them in distinct restricted metastates satisfying~Eq.~(\ref{eq:incong2}). Results from~\cite{ANSW14} and~\cite{ANSW16} then show that the variance of the free energy difference between these two incongruent pure states, whether they appear in the same or different mixed Gibbs state, have free energy fluctuations whose variance scales with the volume, i.e., they satisfy~(\ref{eq:bounds}). The same result applies to the free energy fluctuations of entire mixed Gibbs states at positive temperature as well as ground states at zero temperature, if the edge overlap density of the barycenter~$P_{\rho_J}(q)$ of the corresponding PBC~metastate is a delta-function supported at a single value.

These results provide a proof of a conjecture of Fisher and Huse~\cite{HF87,FH87}, though restricted to the context of finite-volume restrictions of infinite-volume pure states (and ground states under the conditions described in Sect.~\ref{sec:zero}): namely, that the (free) energy difference between incongruent states restricted to a volume~$\Lambda_L$ is typically~O($L^{d/2})$.   However, if one focuses instead on finite systems (as opposed to finite-volume restrictions of infinite systems), an analog of the results presented here provides information on free energy difference fluctuations inside a window far from the boundaries, but not over the entire volume. 

Finally, we showed that the distribution of a partially averaged version of the free energy difference displays Gaussian behavior, in the sense that in the $L\to\infty$ limit this distribution has (at least) a Gaussian tail.  Using this result we obtained results in two dimensions that point toward (but do not yet prove) the existence of a single pure state for all temperatures $T>0$ and a single pair of spin-reversed ground states at zero temperature. 

\medskip

We now turn to a heuristic discussion of the implications of these results (which can be extended to coupling-independent boundary conditions besides periodic, though we shall not pursue that here).  
Our results concern free energy difference fluctuations in finite-volume restrictions of infinite-volume Gibbs states.  An immediate question might be what one expects for analogous quantities in finite-size systems, i.e., a finite box of volume $L^d$ and with periodic or antiperiodic boundary conditions. In this case there is a rigorous {\it upper\/} bound on (free) energy fluctuations between ground states (at zero temperature) or finite-volume Gibbs states (at positive temperature) generated by different boundary conditions: if $F_L^{P}$ denotes the (free) energy inside the volume with periodic boundary conditions and $F_L^{AP}$ is the same quantity but with antiperiodic boundary conditions, then ${\rm Var}(F_L^{P}-F_L^{AP})\le cL^{d-1}$, where $0<c<\infty$ is a constant~\cite{AFunpub,NSunpub,Stein16}. 

For this situation there is no corresponding lower bound (aside from the trivial case of $O(1)$ differences).  As noted above, our results apply to a window far from the boundaries of the finite box, but not over the entire volume. At the same time there is information on these free energy differences over the entire volume owing to multiple numerical studies measuring the spin glass stiffness. These studies measure the free energy difference in volumes using two different boundary conditions (usually but not always periodic and antiperiodic) as a function of size.  Results in three through six dimensions are consistent with free energy fluctuations whose growth is much smaller than linear in the volume~\cite{PY99b,Hartmann99,CBM02,Boettcher04,Boettcher05,WMK14}.  As already emphasized, our results hold only in the infinite-volume setting, so any attempt at comparison with numerics requires a separate analysis.  Here we only briefly note that if one attempts to define spin glass stiffness within a thermodynamic setting, it is difficult in light of the above results to see how to construct a natural definition that doesn't lead to stiffness increasing as the square root of the restricted volume.  If so, this would lead to a disconnect between finite-volume and thermodynamic understandings of spin glass stiffness.  Studies of this are currently ongoing and will be reported elsewhere. 

We now return to the setting of finite-volume restrictions of infinite-volume Gibbs states.  It may seem surprising that free energy difference fluctuations are so large in this setting, especially considering the numerical studies on stiffness mentioned above. Of particular interest is the case of pure states within the same mixed Gibbs states, where free energy differences must be $O(1)$ in the infinite-volume limit.
Then a simple scenario consistent with Theorem~9.1 is that nontrivial mixed Gibbs states do not occur in any finite dimension.  However, they are not altogether ruled out, because these infinite-volume Gibbs states arise from convergent subsequences of finite-volume Gibbs states.  In any such convergent subsequence one takes a fixed window and studies the behavior of the single- and multi-spin correlations inside that window as the overall system size goes to infinity. 

In such a convergent subsequence,  none of the (finite-volume approximations to the) pure states can have free energy differences that scale with the volume, or else they wouldn't appear in the same mixed Gibbs state at all. But given that the free energy difference between pure states $\alpha$ and $\alpha'$ can change sign infinitely often, there could be a {\it sub\/}subsequence of volumes in which the free energy difference between pure states $\alpha$ and $\alpha'$ are $O(1)$, and that subsequence could then converge to a mixed Gibbs state whose decomposition includes $\alpha$ and $\alpha'$.  Of course, one would then require the existence of subsubsubsequences of volumes in which not just two but {\it infinitely many\/} pure states have free energy differences of $O(1)$, because according to Theorem~6.1 a mixed Gibbs state in $\kappa_J$ cannot be decomposed into a finite number (greater than one) of pure state pairs.  Even if this were to occur, it remains the case that any two pure states $\alpha$ and $\alpha'$ appearing in the same mixed state~$\Gamma$ would have large (i.e., order of the square root of the volume) free energy differences inside a fraction close to or equal to one of the restricted volumes~\cite{Readprivate}. 

Our results do not rule out this possibility. But if it were to hold then the (infinite-volume) interface $\alpha\triangle\alpha'$ would have a free energy that scales as $L^{d/2}$ (while changing sign infinitely often). One of the unresolved questions concerning spin glass ground states is a prediction of RSB~\cite{MP01,NS22} asserting the existence of space-filling~(i.e., $d$-dimensional) interfaces (between certain pairs of incongruent ground states) whose energy remains~$O(1)$ on all lengthscales.  If such interfaces did exist, there would, according to Theorems~9.1 and 9.2 and the discussion in this section, be no corresponding space-filling interfaces between {\it pure\/} states at {\it positive\/} temperature whose {\it free\/} energy remains~$O(1)$ on all lengthscales.

\smallskip

{\it Acknowledgments.\/}  We thank Jon Machta and Nick Read for useful discussions, and Nick Read and two anonymous referees for helpful comments on the manuscript.

{\it Data Availability Statement:\/} No datasets were generated or analyzed during the current study.

{\it Competing Interests.\/} The authors have no competing interests to declare that are relevant to the content of this article.  No funding was received for conducting this study.

\bibliography{refs}

\begin{thebibliography}{69}
\bibitem [{ome()}]{omeganote}%
  \BibitemOpen
  \href@noop {} {}\bibinfo {note} {This procedure assumes that in the case
  under consideration the chosen $\omega$ belongs to a single $\Gamma$.
  Otherwise choose a $(\Gamma,\omega)$ pair instead of just
  $\omega$.}\BibitemShut {Stop}%
\end{thebibliography}
\end{document}